\documentclass[showpacs]{revtex4}
\usepackage{amsmath}
\usepackage{amssymb}
\usepackage{epsfig}

\newcommand{\be}{\begin{eqnarray}}
\newcommand{\ee}{\end{eqnarray}}

\begin{document} 
\title{Scattering lengths of strangeness $S=-2$ baryon-baryon interactions}

\author{A.M. Gasparyan$^{1,2}$, J. Haidenbauer$^{3,4}$, and C. Hanhart$^{3,4}$}

\affiliation{
$^1$Institut f\"ur Theoretische Physik II, Ruhr-Universit\"at Bochum, D-44780 Bochum, Germany\\
$^2$SSC RF ITEP, Bolshaya Cheremushkinskaya 25, 117218 Moscow, Russia\\
$^3$Institute for Advanced Simulation, Forschungszentrum J\"ulich, D-52425 J\"ulich, Germany \\
$^4$Institut f\"ur Kernphysik (Theorie) and J\"ulich Center for 
Hadron Physics, Forschungszentrum J\"ulich, D-52425 J\"ulich, Germany
}

\begin{abstract}
We reconsider a method based on dispersion theory, that allows one to extract the 
scattering length of any two-baryon system from corresponding final-state interactions 
in production reactions. 
The application of the method to baryon-baryon systems with strangeness $S=-2$ and $S=-3$ 
systems is discussed. Theoretical uncertainties due to the presence of inelastic channels
with near-by thresholds are examined for the specific situation of the reaction 
$K^- d \to K^0 \Lambda\Lambda$ and the coupling of $\Lambda\Lambda$ to the $\Xi N$ channel. 
The possibility to disentangle spin-triplet and spin-singlet scattering lengths by means of 
various polarization measurements is demonstrated for several production reactions in 
$K^- d$ and $\gamma d$ scattering. 
Employing the method to available data on the $\Lambda\Lambda$ invariant mass 
from the reaction $^{12}C(K^-,K^+\Lambda\Lambda X)$, a $^1S_0$ scattering length of 
$a_{\Lambda\Lambda}=-1.2\pm 0.6$ fm is deduced.
\end{abstract}

\maketitle

\section{Introduction}

The baryon-baryon interaction in the strangeness $S=-2$ sector and, specifically, the
$\Lambda\Lambda$ system has been a topic of interest for quite some time. 
The fascination was generated not least by the possible existence of the so-called 
$H$-dibaryon, a deeply bound 6-quark state with $J=0$, isospin $I=0$, and $S=-2$, 
predicted by R.~Jaffe in 1977 based on a bag-model calculation \cite{Jaffe:1976yi}. 
The binding energies of $\Lambda\Lambda$ in nuclei, deduced from 
sparse information on doubly strange hypernuclei \cite{Dan63,Prowse:1966nz,Aoki:1991ip} 
indicated a strongly attractive ${}^1S_0$ $\Lambda\Lambda$ interaction
and seemed to be at least not inconsistent with the existence of such a bound state. 
The perspective changed drastically when in 2001 a new (and unambiguous)
candidate for ${}_{\Lambda\Lambda}^{\;\;\;6}{\rm He}$ with a much lower 
binding energy was identified \cite{Takahashi:2001nm}, the so-called Nagara event, 
suggesting that the $\Lambda\Lambda$ interaction should be only moderately attractive. 
This conjecture concurs also with evidence provided by various searches for the $H$ 
dibaryon that did not yield any support for its existence, cf. 
Refs.~\cite{Yamamoto00,Merrill01,Yoon07} for the latest experiments. 

However, very recently the $H$-dibaryon was put back on the agenda. Lattice 
QCD calculations by the NPLQCD \cite{Beane10,Beane11a} as well as by the 
HAL QCD \cite{Inoue10,Inoue11}
collaborations provided evidence for a bound $H$-dibaryon. While the actual
computations were performed for pion masses still significantly larger than the
physical one, extrapolations suggest that the $H$-dibaryon could be still 
bound by around 0 -- 7 MeV \cite{Beane11} at the physical point, but it could 
also move above the $\Lambda\Lambda$ threshold and dissolve into the continuum 
\cite{Shanahan11,Hai11}. (The original $H$-dibaryon \cite{Jaffe:1976yi} was 
expected to be bound by roughly 80 MeV!) 

Indeed, the strength of the $\Lambda\Lambda$ interaction as well as those of other 
$S=-2$ baryon-baryon systems is of rather general interest, noteably for a better 
understanding of the role played by the SU(3) flavor symmetry.
Theoretical investigations of the $S=-2$ sector that have been
performed within the conventional meson-exchange picture 
\cite{Stoks99,Rijken06,Rijken06a,Rijken10,Ueda98,Ueda01},
utilizing the constituent quark model \cite{Fujiwara07,Valcarce10},
and also in the framework of chiral effective field theory ($\chi$EFT)
\cite{Polinder07} all rely strongly on the SU(3) symmetry as guideline. 
Furthermore, the hyperon-hyperon ($YY$) interaction plays an important 
role in the understanding of the global properties of compact stars like 
neutron stars. Their stability and size as well as the cooling process
depend sensitively on the strength of the $YY$ interaction 
\cite{Schaffner08,Wang10}. 
 
As indicated above, practically the only experimental constraint we have so far
on the $\Lambda\Lambda$ interaction comes from the analyis of double-$\Lambda$ 
hypernuclei. In the present paper we want to call attention to the fact
that there is also another and even more direct way to determine 
the strength of the $\Lambda\Lambda$ force but also the one in other $S=-2$ systems.
It consists in studying the final-state interaction (FSI) of reactions where 
corresponding pairs of hyperons are produced.  
In fact, recently we proposed a method for extracting hadronic scattering 
lengths from production reactions \cite{Gasparyan:2003cc,Gasparyan:2005fk,Gasparyan:2007mn}.
The presentation of the method in those publications was done with special emphasis on its
application to the hyperon-nucleon ($YN$) interaction. In particular, the reactions
$NN\to KYN$ and $\gamma d\to KYN$ were analyzed, and possible uncertainties of the method were
established. Polarization observables needed to disentangle different spin states of the
final $YN$ system were identified. 

In the present paper we explore the possibility of applying the method proposed 
in \cite{Gasparyan:2003cc,Gasparyan:2005fk,Gasparyan:2007mn} to the $\Lambda\Lambda$ 
system, but also other baryon-baryon states with $S=-2$ or even $S=-3$ are 
considered. 
Our study is motivated by the available data on the $\Lambda\Lambda$
invariant mass distribution determined in the reaction 
$^{12}C(K^-,K^+\Lambda\Lambda X)$ \cite{Ahn98,Yoon07}. 
These data are afflicted by sizeable uncertainties, but still they allow us to 
demonstrate the practicability of our method and to extract an actual value 
for the $\Lambda\Lambda$ ${}^1S_0$ scattering length. 
In order to stimulate future dedicated experiments we consider specifically 
reactions like $K^- d\to K \Lambda\Lambda$ or $K^- d\to K \Xi N$ where
corresponding high-statistics measurements could be performed at J-PARC.
The CLAS collaboration at Jlab has 
measured $\gamma p \to K^+K^+\Xi^-$ \cite{Guo07} and, thus, it might be 
feasible that they can perform also experiments for 
$\gamma d \to K^+K^+\Xi N$ and $\gamma d \to K^+K^0\Lambda\Lambda$
\cite{Schumacher}. Yet another option are reactions like 
$pp\to K^+ K^+ \Lambda\Lambda$ and $pp\to K^0K^0\Sigma^+\Sigma^+$
which could be measured at the future FAIR facility, for example. 
Since the spin structure of these reactions differs partly from the ones 
considered in \cite{Gasparyan:2003cc,Gasparyan:2005fk,Gasparyan:2007mn}
the question of what polarization observables are needed to disentangle the
singlet- and triplet baryon-baryon states has to be re-addressed. 
Note, however, that for the considered $\Lambda\Lambda$ system this issue is 
not relevant. Near threshold it can only be in the spin-singlet ($^1S_0$) 
state. 
Due to the Pauli principle the other $S$-wave, the $^3S_1$, is 
forbidden in this case. Thus, no polarization experiment is 
required and, consequently, our method could be even applied to 
data on $\Lambda\Lambda$ production on somewhat heavier nuclei,
e.g. in $K^-\,^3{\rm He}$ or $K^-\,^4{\rm He}$.

Independently of that, the error estimation \cite{Gasparyan:2003cc} for 
the method has to be re-done. 
Specifically, for the $\Lambda\Lambda$ system the inelastic threshold 
(due to the $\Xi N$ channel) lies with around 25 MeV much lower than 
for $\Lambda N$ (where it is around 80 MeV and due to the $\Sigma N$ channel). 
However, as we will see, the latter aspect increases the 
theoretical error of our method only marginally, under the discussed
reasonable assumptions, and in the absence of a bound state. Taking into
account the possibility of a bound state requires much more effort and is 
technically more complicated. In view of the current extrapolations of the
lattice QCD results which rather seem to disfavor the existence of a bound state 
\cite{Beane11,Shanahan11,Hai11} we avoid the pertinent complications 
in the present study. 

For completeness let us mention that FSI effects as a tool to constrain
the $\Lambda\Lambda$ interaction were considered already many years
ago by Afnan and collaborators \cite{Afnan97,Carr98}. Their study
was done under rather different presuppositions, namely for the
reaction $\Xi^- d \to n \Lambda\Lambda$ and within the framework
of Faddeev equations. With regard to the $\Xi N$ system 
there is also an entirely different possibility to determine the 
corresponding scattering lengths, namely via the study of
$\Xi^-$ atoms. Shifts of the energy levels due to the presence of the 
strong interaction would permit to deduce the scattering lengths
for $\Xi^-p$ or $\Xi^-d$, say, via the Deser-Trueman formula. 
The prospects of corresponding experiments were discussed in
Ref.~\cite{Batty99}. 

The paper is structured as follows. 
In Sect. II we review briefly our method. Section III is devoted
to the $\Lambda\Lambda$ system. First we provide a new estimation 
for the error of the scattering length due to the extraction method, 
taking into account the relatively small separation of the 
$\Lambda\Lambda$ and $\Xi N$ thresholds. 
Then we apply our method to available data on the $\Lambda\Lambda$ 
invariant mass spectrum from a measurement of the reaction
$^{12}C(K^-,K^+\Lambda\Lambda X)$. 
In Sect. IV we discuss several aspects of applying our method also
to the $\Xi N$ and $\Sigma\Sigma$ final-state interactions and even
to the strangeness $S=-3$ sector. 
The paper ends with a short summary. 
Details of the polarization observables required for separating the
spin singlet and triplet states are summarized in an appendix. 

\section{Review of the method}
\label{method}
The basic idea of the method is to exploit the scale separation between a short-ranged production operator
and a long-ranged final-state interaction (FSI). In this case the production operator can be regarded as
point-like, and the FSI can be factored out. These conditions restrict the class of reactions and 
kinematic regimes that one can consider.
Namely, one can only apply the method to reactions with large momentum transfer $q_t$. 
Furthermore, the scattering length $a$ in the system under consideration must have an
appropriate magnitude, i.e. fulfil the condition $a\gg 1/q_t$. Sufficiently large 
scattering lengths are expected in the baryon-baryon sector. 
In particular, it is interesting to study the hyperon-nucleon and hyperon-hyperon interactions 
with different strangeness content of the hyperons. An elegant way to utilize the condition of 
scale separation is a dispersion-relation approach. Imposing unitarity and analyticity constraints 
on the amplitude and assuming that there are no bound states, one arrives at the following expression
for the reaction amplitude $A_S$ \cite{Muskhelishvili1953,Omnes:1958hv,Gasparyan:2003cc}
\begin{eqnarray}
A_S(s,t,m^2)=\exp\left[{\frac1\pi\int_{m_{0}^2}^{m_{max}^2}\frac{\delta_S(m'^2)}{m'^2-m^2-i0}dm'^2}\right]\Phi(s,t,m^2),
\label{OM}
\end{eqnarray}
where $m$ is the invariant mass of the produced baryon-baryon system with the threshold value $m_0$, 
$s$ is the total center-of-mass (CM) energy squared, and $t$ represents all the remaining 
kinematic variables the amplitude depends upon. 
The function $\Phi(s,t,m^2)$ slowly varies with $m^2$, which is a consequence of 
the assumed large momentum transfer. The cut off $m_{max}$ has to be determined in such a way that 
the integral extends over the whole region where FSI effects are expected to be important. 
Based on scale arguments a condition for $m_{max}$ was derived in Ref.~\cite{Gasparyan:2003cc}
which reads, re-formulated in terms of the maximum kinetic energy in the two-baryon system,
$\epsilon_{max}=m_{max}-m_0 \gtrsim \frac{1}{2 a_S^2 \mu}$.
Here $a_S$ is the scattering length in question and $\mu$ is the reduced mass of the
baryon-baryon system. 
As argued in Ref.~\cite{Gasparyan:2003cc}, for the hyperon-nucleon interaction a typical 
cut off is given by the condition $\epsilon_{max}\approx 40$ MeV.
The baryon-baryon scattering process has to be elastic in this region (i.e. there should be no 
other open channels) and it should be dominated by the $s$-wave amplitude
parametrized by the phase shift $\delta_S(m^2)$. Note that formula (\ref{OM}) can only be applied 
to amplitudes for a specific baryon-baryon spin state $S$. Therefore, one has to be able to 
separate spin-singlet and spin-triplet states experimentally. The index ``S'' on the quantities
above (and below) is a reminder that one has to consider the production of the 
baryon-baryon system in a definite spin $S$. 

It was shown in \cite{Gasparyan:2003cc} how one can invert Eq.~(\ref{OM}) to express the scattering
length via the reaction amplitude squared (or the differential cross section
$\frac{d^2\sigma_S}{dm'^2dt}$) 
\begin{eqnarray}
 \nonumber a_S&=&\lim_{{m}^2\to m^2_{0}}\frac1{2\pi}\left(\frac{m_a+m_b}
{\sqrt{m_a m_b}}\right){\bf P}
\int_{m^2_{0}}^{m^2_{max}}dm'^2\sqrt{\frac{m^2_{max}-{m}^2}{m^2_{max}-m'^2}}\\
& \times& \ \frac1{\sqrt{m'^2-m^2_{0}}\,(m'^2-{m}^2)}
\log{\left\{\frac{1}{p'}\left(\frac{d^2\sigma_S}{dm'^2dt}\right)\right\}},
\label{a}
\end{eqnarray}
where $m_a$ and $m_b$ are the masses of the two baryons, $m_0=m_a+m_b$, 
$p'$ is the CM momentum in the baryon-baryon system and {\bf P} indicates that
the principal value of the intregral has to be taken. 
An analogous equation can be derived for the effective range.

Possible theoretical uncertainties of the method originate from the following sources:
(i) energy dependence of the production operator,
(ii) influence of scattering at higher energies ($m>m_{max}$),
(iii) contributions from inelastic channels (e.g. from the $\Sigma N\leftrightarrow\Lambda N$ transition) and
(iv) final state interaction among other pairs of particles.
For the hyperon-nucleon FSI the theoretical uncertainty in the determination of the scattering
length was estimated to be $0.3$ fm at most \cite{Gasparyan:2003cc}. This estimate was
confirmed by model calculations of production amplitudes using several different models
for the hyperon-nucleon interactions with triplet and singlet scattering
lengths varying from $-0.7$ to $-2.5$ fm. 

The general form of Eq.~(\ref{OM}) admits approximations under certain conditions.
One of the standard approximative treatments follows from the assumption that the
phase shifts are given by the first two terms in the effective range
expansion,
\begin{eqnarray}
p \ {\rm cot} (\delta(m^2)) = -{1 \over a}+{r_e\over 2} p \,^2 \ ,
\label{ere}
\end{eqnarray}
over the whole energy range, which is usually called the effective range approximation (ERA). 
In this case the relevant integrals (\ref{OM}) can be evaluated in closed
form as \cite{Goldberger1964}
\begin{eqnarray}
A(m^2) \propto
\frac{(p^2+\alpha^2)r_e/2}{-1/a+(r_e/2)p^2-ip} \ ,
\label{arform}
\end{eqnarray}
where $\alpha = 1/r_e(1+\sqrt{1-2r_e/a})$.
Because of its simplicity Eq. (\ref{arform}) is often used for the treatment
of the FSI.
A further simplification can be made if one assumes that $a\gg r_e$, a situation that 
is realized in the $^1S_0$ partial wave of the $NN$ system. Then
the energy dependence of the quantity in Eq. (\ref{arform})
is given by  the energy dependence of the elastic  amplitude
\begin{eqnarray}
A(m^2) \propto
\frac{1}{-1/a+(r_e/2)p^2-ip} \ ,
\label{MW}
\end{eqnarray}
as long as $p \ll 1/r_e$.
Therefore one expects that, at least for
small kinetic energies, $NN$ elastic scattering and particle production reactions with 
a $NN$ final state exhibit the same energy dependence 
\cite{Goldberger1964,Watson:1952ji,Migdal:1966tq,Baru00},
which indeed was experimentally confirmed for meson production \cite{Hanhart}. 
The treatment of FSI effects based on Eq.~(\ref{MW}) is often referred to as 
Migdal-Watson approach \cite{Watson:1952ji,Migdal:1966tq}, the one utilizing
Eq.~(\ref{arform}) as Jost-function approach. 
The reliability of such approximations as compared to the formula~(\ref{a})
was investigated in detail in \cite{Gasparyan:2005fk}.
In general the method based on Eq.~(\ref{a}) works systematically better than the approximations 
and gives scattering lengths within $0.3$ fm accuracy even for rather large 
scattering lengths like those for $NN$ scattering.
The uncertainty in the extraction employing the other two methods is typically larger.
As demonstrated in \cite{Gasparyan:2005fk} these procedure lead to a systematic 
deviation from the true values of the scattering lengths of the order of 0.3 fm
(Jost) and of 0.7 fm (Migdal-Watson).

\begin{figure}[h]
\begin{center}
\epsfig{file=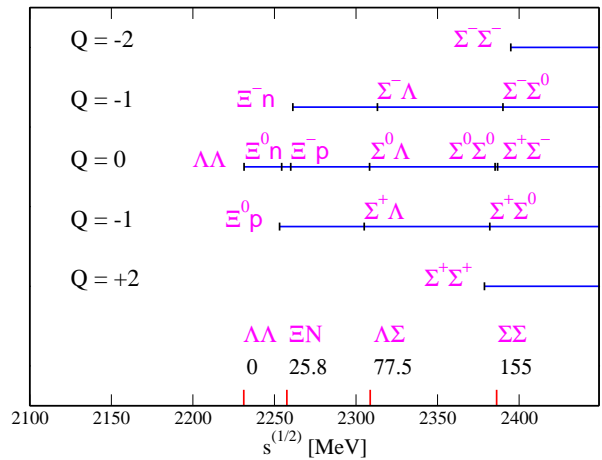, width=6.0cm,angle=-90}
\end{center}
\caption{Graphic representation of the thresholds for the various
strangeness $S=-2$ channels with charges from $Q=-2$ to $Q=+2$. 
The thresholds for isospin averaged masses are indicated at the
bottom. 
}
\label{fig0}
\end{figure}

\section{The $\Lambda\Lambda$ scattering length}

The $\Lambda\Lambda$ system is certainly the most promising case where one
could apply our method. Its threshold is the lowest one among all $S=-2$ 
channels and measurements could be performed for the reaction 
$K^-d \to K^0 \Lambda \Lambda$, for example. Moreover, no polarization experiment is
required because (near threshold) the $\Lambda\Lambda$ can be only in the 
(spin singlet) $^1S_0$ partial wave. The $^3S_1$ state is forbidden 
due to the Pauli principle, as already mentioned. Thus, spin triplet states 
can only occur in $P$ (or higher partial) waves - and it is save to assume 
that such higher partial waves do not contribute near threshold.
There is, however, a complication because the first inelastic threshold 
(due to $\Xi N$) is fairly close: the $\Xi^0n$ channel opens at an 
excess energy of 23.06 MeV, c.f. Fig.~\ref{fig0}. 
(In the $\Lambda p$ case considered in \cite{Gasparyan:2003cc} the first 
inelastic channel ($\Sigma^0 p$) opens at 76.96 MeV!) Thus, it is necessary 
to re-do the error estimate of Ref. \cite{Gasparyan:2003cc}. This will be done 
in subsection A below. Note that, for convenience, we will work with 
isospin-averaged masses throughout this section so that the $\Xi N$ 
threshold is located at 25.8 MeV! 
 
In subsection B we apply our method to the $\Lambda\Lambda$ invariant mass 
distribution measured in the reaction $^{12}C(K^-,K^+\Lambda\Lambda X)$ 
by the KEK-PS E224 
Collaboration \cite{Yoon07}. Those data, though afflicted by sizeable
error bars, allow us to demonstrate how our method works, and they even enable
us to deduce a concrete value for the $\Lambda\Lambda$ $^1S_0$ scattering 
length. 

\subsection{Error estimation}
In this subsection, we generalize the discussion of theoretical errors presented 
in \cite{Gasparyan:2003cc} to the case of the occurrence of inelastic channels. 
Specifically, 
we estimate the theoretical error for the extraction of the hyperon-hyperon 
scattering length exemplary for the reaction $K^-d \to K^0 \Lambda \Lambda$ 
taking into account that there is a near-by inelastic threshold due to the
coupling of $\Lambda \Lambda$ to the $\Xi N$ channel. 
 
The uncertainties originate \cite{Gasparyan:2003cc} from the energy (i.e. $m^2$) 
dependence of the function $\Phi(s,t,m^2)$ in Eq.~(\ref{OM}).
These include the energy dependence of the production operator (i.e. the influence of left-hand singularities), 
contributions of the elastic scattering to the dispersion integral at higher energies, 
the influence of inelastic 
channels, and the interaction between other pairs of particles in the final state.
The latter effect can be controlled by choosing different kinematical conditions such as initial 
energy (final-state interaction among other pairs of particles would depend on such a choice whereas the 
$\Lambda\Lambda$ FSI does not).
Also investigating the invariant-mass distribution for a corresponding pair of particles 
via a Dalitz plot analysis can provide additional information on their interaction \cite{Hanhart}. 
In what follows, we will disregard this (possibly important) kind of correction and focus on the other three.

The energy dependence of the production amplitude $A_S(s,t,m^2)$ 
can be deduced from the basic principles of analyticity and unitarity.
The discontinuity of the amplitude in $m^2$ is given by the sum of the elastic term, the inelastic 
contribution of the $\Xi N$ (and/or other) channel, $D_S^{in}(s,t,m^2)$, and the left-hand part,
$D_S^{l.h.}(s,t,m^2)$, denoting the remaining contribution from the production operator
\begin{eqnarray}
D_S(s,t,m^2)\equiv \frac{1}{2i}(A_S(s,t,m^2+i0)-A_S(s,t,m^2-i0))=A(s,t,m^2)e^{-i\delta}\sin{\delta}+D_S^{l.h.}(s,t,m^2)+D_S^{in}(s,t,m^2).
\label{D_S}
\end{eqnarray}
The inelastic contribution reads
\begin{eqnarray}
D_S^{in}(s,t,m^2)&=&\left(\frac{A(s,t,m^2)(1-\eta)e^{-2i\delta}}{2i}+A_2(s,t,m^2)f_{12}^*(m^2)p_2\right)\theta(m^2-m_2^2), 
\ m_2^2=(m_\Xi+m_N)^2  ,\nonumber\\
p_2&=&\sqrt{\frac{(m^2-m_2^2)m_\Xi m_N}{m_2^2}}\approx \frac{1}{2}\sqrt{m^2-m_2^2} \ \ \text{(in non-relativistic kinematics)},
\label{D_in}
\end{eqnarray}
where $\eta$ is the inelasticity parameter in the $\Lambda\Lambda$ system, $A_2$ is 
the production amplitude for the reaction $K^-d\to K^+\Xi N$, and $f_{12}$ is the 
$\Lambda\Lambda\to\Xi N$ transition amplitude. The latter can be written in terms of $\eta$, $\delta$, 
and the phase shift $\delta_2$ of the $\Xi N$ channel:
\begin{eqnarray}
f_{12}=\frac{\sqrt{1-\eta^2}\, e^{i(\delta+\delta_2)}}{2\sqrt{p \,p_2}} \ .
\end{eqnarray}
In order to shorten the notation we rewrite $D_S^{in}$  as
\begin{eqnarray}
D_S^{in}(m^2)&=&A(m^2)\theta(m^2-m_2^2)\left(\frac{1-\eta}{2i}e^{-2i\delta}+\left|\frac{A_2(m^2)}{A(m^2)}\right|\frac{|\tilde f_{12}(m^2)|}{2}e^{-i\tilde\delta}\right) \,,\nonumber\\
\tilde f_{12}(m^2)&=&2 p_2 f_{12}(m^2) \,,\ \tilde\delta=\delta+\delta_2+\delta_{A}-\delta_{A_2}\,,
\label{D_in2}
\end{eqnarray}
where we suppressed any dependence on $s$ and $t$, and where we denoted the phases 
of the production amplitudes $A$ and $A_2$ by $\delta_{A}$ and $\delta_{A_2}$, respectively.
The solution of Eq.~(\ref{D_S}) in the 
physical region can be represented as (see Refs. \cite{Muskhelishvili1953,Omnes:1958hv,Frazer1959,Gasparyan:2003cc}
\begin{eqnarray}
A(m^2)&=&e^{\displaystyle{u(m^2)}}\tilde\Phi(m^2)\equiv e^{\displaystyle{u(m^2)}}\left(\Phi_{l.h.}(m^2)+\Phi_{in}(m^2)\right),\nonumber\\
\Phi_{l.h.}(m^2)&=&\int_{-\infty}^{\tilde m^2} \frac{ D_S^{l.h.}(m' \, ^2)e^{\displaystyle{-u(m' \, ^2)}}}{m' \, ^2-m^2-i0}\frac{dm' \, ^2}{\pi}\,,\ 
\Phi_{in}(m^2)=\int_{m_2^2}^{+\infty} \frac{ D_S^{in}(m' \, ^2)e^{\displaystyle{-u(m' \, ^2)}}(m^2-m_0^2)}{(m' \, ^2-m^2-i0)(m' \, ^2-m_0^2)}\frac{dm' \, ^2}{\pi},\nonumber\\
u(m^2)&=&\frac1\pi\int_{m_0^2}^\infty\frac{\delta(m' \, ^2)(m^2-m_0^2)}{(m' \, ^2-m^2-i0)(m' \, ^2-m_0^2)}dm' \, ^2.
\label{Frazer}
\end{eqnarray}
where $\tilde m^2$ denotes the upper end of the left-hand cut. 
In order to 
remove the energy independent part from  the inelastic dispersion integral, we made subtractions at 
$m^2=m_0^2$ in the definition of $u(m^2)$ and in the inelastic dispersion integral (a constant term is 
assigned then to the left-hand contribution which anyway, as we will show, is slowly varying with energy.)

The theoretical error of the extracted scattering length is determined by the energy dependence 
of the function $\Phi(s,t,m^2)$ from Eq.~(\ref{OM}) \cite{Gasparyan:2003cc}
\begin{eqnarray}
\delta a^{(th)}= -\lim_{{m}^2\to
m_0^2}{\bf P}
\int_{m_0^2}^{m_{max}^2}\frac{\log{| \Phi(m' \, ^2)|^2}}
{\sqrt{m' \, ^2-m_0^2}\, (m' \, ^2-m^2)}\sqrt{\frac{m_{max}^2-m^2}{m_{max}^2-m'
\, ^2}}\frac{dm' \, ^2}{\pi}.
\label{deltaath}
\end{eqnarray} 
The function $\Phi(m^2)$ depends on $m_{max}^2$. This dependence factors out  \cite{Gasparyan:2003cc}
$\Phi(m_{max}^2,m^2)=\Psi(m_{max}^2,m^2)\tilde\Phi(m^2)$ where $\Psi(m_{max}^2,m^2)$ contains the 
information on the phase shift at energies above $m_{max}$
\begin{eqnarray}
 \Psi(m_{max}^2,m^2)\propto\exp\left[{\frac1\pi\int_{m_{max}^2}^{\infty}\frac{\delta(m' \, ^2)}{m' \, ^2-m^2-i0}dm' \, ^2}\right].
\label{dis3}
\end{eqnarray}
Thus the theoretical error is the sum  
$\delta a^{(th)}=\delta a^{m_{max}}+\delta \tilde a$, where $\delta a^{m_{max}}$ is due to 
the factor $\Psi(m_{max}^2,m^2)$ and $\delta \tilde a$ is determined by the energy 
dependence of $\tilde\Phi(m^2)$ which is related to the energy dependence of the production 
operator and inelastic effects.
 
\begin{figure}[h]
\includegraphics*[width=6cm,height=3cm]{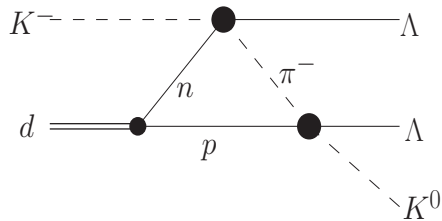}
\caption{Typical production mechanism for the reaction $K^- d\to K^0 \Lambda\Lambda$. }
\label{diagram}
\end{figure}

Let us first estimate the theoretical error originating from the energy dependence of the production 
operator, neglecting for the moment the inelastic contributions. In order to do this one has to 
investigate the contribution to the dispersion integral from the left-hand cuts. We follow here 
the procedure utilized in \cite{Gasparyan:2003cc}.
We are interested in the left-hand singularities of the amplitude, i.e. singularities in some 
momentum transfer variable $t$. The simplest production mechanism for the reaction 
$K^- d\to K^0 \Lambda\Lambda$ is seemingly the one shown in Fig.~\ref{diagram} denoting 
the exchange of one nucleon and one pion (one should add, of course, the diagram with proton 
and nucleon interchanged and $\pi^+$ replaced by $\pi^0$). 
Clearly, there are also other more complicated production mechanisms that will
contribute. However, those should be of even shorter range and thus, correspond to 
production operators with even weaker energy dependence so that they are not relevant
for the estimation of the theoretical error. 
We consider for simplicity kinematics corresponding to the final particles produced not far away 
from threshold (so that the momenta of the final particles are small compared to the momentum transfer). 
The required initial CM momentum is then $p_i\approx 580$ MeV/c. The production 
operator contains a cut corresponding to the $\pi^- p$ intermediate state, which can be associated 
with the interaction in the $\Lambda K^0$ system discussed above. The other cut over the neutron 
and $\pi^-$ is the one we are interested in. We can very roughly estimate the energy dependence 
associated with this singularity via approximating it by a pole term,
\begin{eqnarray}
\Phi(m^2)\sim\frac{1}{t-m_{n\pi^-}^2}, \ t=(p_{K^-}-p_\Lambda)^2 \ ,
\end{eqnarray}
where $p_{K^-}$ and $p_\Lambda$ are the corresponding 4-momenta and $m_{n\pi^-}$ is the 
effective invariant mass of the $n\pi^-$ system, i.e. $m_{n\pi^-}\approx 1$ GeV.
For threshold kinematics (i.e. zero momentum of all final particles -- 
this choice is made for definiteness) 
this energy dependence (after averaging over all directions) has the form 
$\Phi\approx 1+ p^2/p_t^2$, where $p$ is 
the CM momentum in the $\Lambda\Lambda$ system and $p_t$ is of the order of $2$ GeV. 
The correction to the scattering length due to such an energy dependence of the amplitude 
amounts to \cite{Gasparyan:2003cc} $\frac{p_{max}}{p_t^2}\approx 0.01$ fm for values of 
$\epsilon_{max}=40-60$ MeV. The above rough estimation is sufficient to observe that the 
error coming from the energy dependence of the production operator is negligible as 
compared to the other sources of uncertainties and we can savely assume that 
\begin{eqnarray}
\Phi_{l.h.}(m^2)\approx\Phi_{l.h.}(m_0^2)=A(m_0^2)\approx A(m^2)e^{\displaystyle{-u(m^2)}}.
\label{l.h.}
 \end{eqnarray}
Note that since the $\Lambda\Lambda$ scattering length is expected to be somewhat 
smaller than $a_{\Lambda p}$ \cite{Stoks99,Fujiwara07,Polinder07,Rijken10} 
we choose  $\epsilon_{max}=60$ MeV rather than $\epsilon_{max}=40$ MeV (used for $\Lambda N$ scattering) 
as our central value in order to minimize the effect of higher energy scattering. 
As was pointed out in Ref.~\cite{Gasparyan:2003cc} $\epsilon_{max}$ must be chosen 
well above $\frac{1}{a_S^2 m_\Lambda}$ (cf. also the discussion in Sect.~\ref{method}).
Although a further increase of $\epsilon_{max}$ 
can help even more in reducing this effect, in reality it would be difficult in this case 
to separate $S$-waves from higher partial waves in the final state and to avoid the influence 
of the interaction in other channels -- the effects and the number of 
inelastic channels would increase.

Next we consider the error coming from the inelastic channels coupled to the 
$\Lambda\Lambda$ system. For simplicity we consider only the one nearest to the 
$\Lambda\Lambda$ threshold, namely $\Xi N$, which opens only about $25$ MeV 
above the $\Lambda\Lambda$ threshold so that it is necessary to analyze its impact.
Clearly, due to the lack of empirical information on the $\Xi N$ interaction and 
the coupling of this channel to the $\Lambda\Lambda$ system 
such an error analysis cannot be done in a completely model-independent way. 
One has to make some assumptions on the strength of the interactions in the
relevant channels.  
For our analysis we prepared three variants of the hyperon-hyperon 
interaction of \cite{Polinder07} which yield $\Lambda \Lambda$ 
scattering lengths of $-1.36$ fm, $-1.50$ fm, and $-1.70$ fm, respectively. 
We are interested in the situation when the effect coming from the inelastic 
channel is small and its contribution can be treated perturbatively, i.e.
\begin{eqnarray} 
\Phi_{in}(m\, ^2)\ll\Phi_{l.h.}(m\, ^2) .  
\label{perturbativity}
\end{eqnarray}
If this is not the case the error of the extraction of the scattering length would 
be comparable with the scattering length itself and then an extraction 
would no longer be meaningful. We made a subtraction in Eq.~(\ref{Frazer}) at $m^2=m_0^2$ 
so that $\Phi_{in}$ is exactly zero at the beginning 
of the integration interval. Then a small resulting correction to the 
scattering length would imply a likewise small variation of $\Phi_{in}$ 
so that then the condition~(\ref{perturbativity}) would be justified.
Note that formally the 
subtraction point does not enter the expression for the error.

Utilizing Eqs.~(\ref{perturbativity}) and (\ref{l.h.}) one can rewrite the formula (\ref{deltaath}) 
for the inelastic contribution to the theoretical error in the form

\begin{eqnarray}
\delta a^{in}&=& -\lim_{{m}^2\to
m_0^2}\frac1{\pi} {\bf P}
\int_{m_0^2}^{m_{max}^2}\frac{\log{| \Phi_{l.h.}(m' \, ^2)+\Phi_{in}(m' \, ^2)|^2}}
{\sqrt{m' \, ^2-m_0^2}\,(m' \, ^2-m^2)}\sqrt{\frac{m_{max}^2-m^2}{m_{max}^2-m'
\, ^2}}dm' \, ^2 \nonumber\\
&\approx&-\lim_{{m}^2\to
m_0^2}\frac1{\pi} {\bf P}
\int_{m_0^2}^{m_{max}^2}\frac{2\Re\{\Phi_{in}(m' \, ^2)/A(m_0^2)\}}
{\sqrt{m' \, ^2-m_0^2}\,(m' \, ^2-m^2)}\sqrt{\frac{m_{max}^2-m^2}{m_{max}^2-m'
\, ^2}}dm' \, ^2 \nonumber\\
&=&-\frac1{\pi} 
\int_{m_0^2}^{m_{max}^2}\frac{2\Re\{(\Phi_{in}(m' \, ^2)-\Phi_{in}(m_0^2))/A(m_0^2)\}}
{\sqrt{m' \, ^2-m_0^2}\,(m' \, ^2-m_0^2)}\sqrt{\frac{m_{max}^2-m_0^2}{m_{max}^2-m'
\, ^2}}dm' \, ^2 .
\label{delta_a_in}
\end{eqnarray} 
Substituting $\Phi_{in}(m^2)$ from Eq.~(\ref{Frazer}) and performing one integration one gets
\begin{eqnarray}
\delta a^{in}&=&\Im\frac{4p_{max}}{\pi} \int_{m_2^2}^{m_{max}^2}dm^2 \frac{D_S^{in}(m^2)e^{\displaystyle{-u(m^2)}}}{A(m_0^2)(m^2-m_0^2)^{\frac{3}{2}}
\sqrt{m_{max}^2-m^2}}
 \nonumber\\
&-&\Re\frac{4p_{max}}{\pi} \int_{m_{max}^2}^{+\infty}dm^2 \frac{D_S^{in}(m^2)e^{\displaystyle{-u(m^2)}}}{A(m_0^2)(m^2-m_0^2)^{\frac{3}{2}}
\sqrt{m^2-m_{max}^2}},
\label{delta_a_in2}
\end{eqnarray} 
where $p_{max}=\frac{1}{2}\sqrt{m_{max}^2-m_0^2}$.

Using the explicit form of $D_S^{in}(m^2)$ from Eq.~(\ref{D_in}) and making a suitable change of variables we have
\begin{eqnarray}
\delta a^{in}&=&\delta a^{in}_1+\delta a^{in}_2\,,\nonumber\\
\delta a^{in}_1&=&\frac{1}{\pi p_{max}} \left(\int_0^{y_2}
 \frac{d\tilde y}{(1-\tilde y^2)^{(3/2)}} (\eta-1) \cos{(2\delta)}+ \int_0^\infty \frac{dy}{(1+y^2)^{(3/2)}}(1-\eta) \sin{(2\delta)}
\right)\,,
 \nonumber\\
\delta a^{in}_2&=&-\frac{1}{\pi p_{max}}\left(\int_0^{y_2} \frac{d\tilde y}{(1-\tilde y^2)^{(3/2)}}|\tilde f_{12}| \sin{\tilde\delta}\left|\frac{A_2}{A}\right|
+ \int_0^\infty \frac{dy}{(1+y^2)^{(3/2)}} |\tilde f_{12}| \cos{\tilde\delta}\left|\frac{A_2}{A}\right|
\right)
 \,,
\end{eqnarray}
with $\tilde y=\sqrt{\frac{m_{max}^2-m^2}{m_{max}^2-m_0^2}}$, 
$y_2=\sqrt{\frac{m_{max}^2-m_2^2}{m_{max}^2-m_0^2}}$, and 
$y=\sqrt{\frac{m^2-m_{max}^2}{m_{max}^2-m_0^2}}$.
From \cite{Gasparyan:2003cc} we recall the expression for $\delta a^{m_{max}}$ (with the same definition of $y$)
\begin{eqnarray}
 \delta a^{m_{max}}=\frac2{\pi p_{max}}\int_0^\infty
 \frac{\delta(y)dy}{(1+y^2)^{(3/2)}}. 
\end{eqnarray}
Note that the integration over $y$ from $0$ to $\infty$ can be truncated at $y=1$, say, 
for practical applications since the integrals are rapidly converging (unless the phase shift 
is rising unnaturally fast with energy). In any case, we cannot trust the $\chi$EFT 
predictions for the amplitude at such high energies.

The numerical values of $\tilde\delta$ and of $\left|\frac{A_2}{A}\right|$ are not known, therefore we estimate
\begin{eqnarray}
|\delta a^{in}_2|&<&\frac{1}{\pi p_{max}}\left(\int_0^{y_2} \frac{d\tilde y}{(1-\tilde y^2)^{(3/2)}}|\tilde f_{12}| 
+ \int_0^\infty \frac{dy}{(1+y^2)^{(3/2)}} |\tilde f_{12}|
\right){\rm Max}\left|\frac{A_2}{A}\right|
 \,,
\end{eqnarray}
where ${\rm Max}\left|\frac{A_2}{A}\right|$ is the maximal value of this ratio in the considered energy region.

Using the three mentioned variants of the hyperon-hyperon interaction we arrive at the 
following estimates of the theoretical errors for $\epsilon_{max}=60$ MeV:
\begin{eqnarray}
\delta a^{in}_1+\delta a^{m_{max}}=-0.19 \ {\rm fm}\,,|\delta a^{in}_2|<0.28 \,{\rm Max} \left|\frac{A_2}{A}\right|{\rm fm}\,,\nonumber
\end{eqnarray}
for the variant with $a=-1.36$ fm,
\begin{eqnarray}
\delta a^{in}_1+\delta a^{m_{max}}=-0.11 \ {\rm fm}\,,|\delta a^{in}_2|<0.30 \,{\rm Max} \left|\frac{A_2}{A}\right|{\rm fm}\,,\nonumber
\end{eqnarray}
for the variant with $a=-1.50$ fm,
\begin{eqnarray}
\delta a^{in}_1+\delta a^{m_{max}}=-0.22 \ {\rm fm}\,,|\delta a^{in}_2|<0.14 \,{\rm Max} \left|\frac{A_2}{A}\right|{\rm fm}\,,\nonumber
\end{eqnarray}
for the variant with $a=-1.70$ fm.
The value of ${\rm Max}\left|\frac{A_2}{A}\right|$ can only be accessed from the 
corresponding production experiment for the $\Xi N$ channel. Some estimates can be 
obtained by looking at the strength of the cusp effect in the $\Lambda\Lambda$ 
production channel. Under the assumption that there is no specific production 
mechanism that makes this ratio large, one can estimate the ratio, at least 
qualitatively, from the unitarity contribution that correspond to the (on shell) 
$\Lambda\Lambda\to\Xi N$ conversion. 
It amounts to ${\rm Max}\left|\frac{A_2}{A}\right|={\rm Max}|f_{12}p_2|<0.3$ 
for all three considered interactions.
Based on those numbers a rough estimation for the full theoretical error related to 
inelastic effects and higher energy scattering yields $\delta a^{th}<0.3-0.4$ fm.
Such small values for $\delta a^{th}$ justify the approximation made in 
Eq.~(\ref{perturbativity}), because it means that $\Phi_{in}(m\, ^2)$ does not 
change much for $p<\frac{1}{\delta a^{th}}$, an energy range that savely covers 
the region we are interested in.

\begin{figure}[t]
 \includegraphics*[width=7cm]{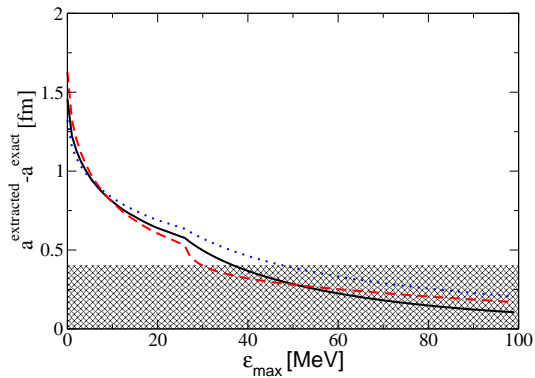}
 \caption{Dependence of the extracted scattering lengths
on the value of the upper limit of integration, $\epsilon_{max}$.
Shown is the difference to the exact results for three variants of the $\chi$EFT
$\Lambda\Lambda$ interaction with $a=-1.50$ fm (solid line), $a=-1.70$ fm (dashed line), 
and $a=-1.36$ fm (dotted line).
The shaded area indicates the estimated error of the applied method.
}
 \label{delta_a}
 \end{figure}

In order to test our method and to check our error analysis we applied it to 
a production amplitude, calculated with a point-like production operator,
that incorporates $\Lambda\Lambda$ final-state interactions generated from 
the three variants of the $\chi$EFT interaction. 
In Fig.~\ref{delta_a} we show the dependence of the difference 
of the extracted scattering length and the exact one 
on the cut off $\epsilon_{max}$ of the integration. 
One can see that for $\epsilon_{max}=60$ MeV the 
theoretical error is indeed within the range of $0.25\,-\,0.35$ fm in agreement 
with the preceding analysis. Note the apparent drop of the curves around 25 MeV, 
i.e. at the opening of the $\Xi N$ channel. 

\subsection{Analysis of data on the $\Lambda\Lambda$ invariant mass from 
$^{12}C(K^-,K^+\Lambda\Lambda X)$}

First results for the $\Lambda\Lambda$ invariant mass distribution 
were reported by the KEK-PS E224 Collaboration from a measurement
of the $^{12}C(K^-,K^+\Lambda\Lambda X)$ reaction in 1998 \cite{Ahn98}. An enhancement
was seen for invariant masses near threshold. Already at that time
there were attempts to extract the $\Lambda\Lambda$ interaction from
the spectrum \cite{Ohnishi00,Ohnishi01a}. Then in 2007
the KEK-PS E522 Collaboration published a $\Lambda\Lambda$ invariant 
mass spectrum with somewhat better statistics \cite{Yoon07}. Also in 
this case efforts were made to extract the $\Lambda\Lambda$ scattering 
length. The value reported at some conferences \cite{Yoon08,Yoon10}, 
$a_{\Lambda\Lambda} = -0.10^{+0.37}_{-1.56} \pm 0.28$ fm,
was obtained by utilizing the Migdal-Watson approach (Eq.~(\ref{MW})). 

\begin{figure}[t]
\includegraphics*[width=7cm]{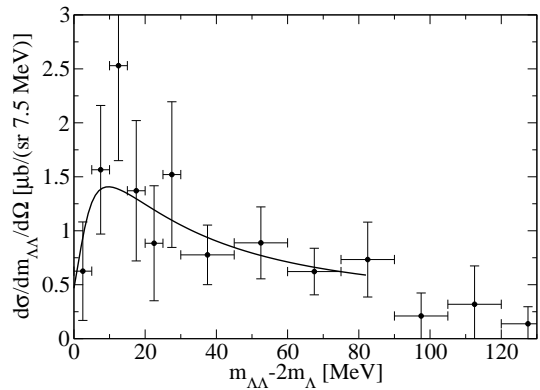}
\caption{$\Lambda\Lambda$ invariant mass spectrum for the reaction  
$^{12}C(K^-,K^+  \Lambda\Lambda X)$ \cite{Yoon07} and our fit to it (solid line)}.
\label{Yoon}
\end{figure}

As already pointed out above and as we thoroughly investigated in
\cite{Gasparyan:2005fk}, the Migdal-Watson approach works only well 
for fairly large scattering lengths, 
i.e. for values of the order of $5$ fm or more,
as they are typical for the $NN$ interaction. For small scattering
lengths as suggested by the analyis in \cite{Yoon08,Yoon10} this
approach is not reliable. It can lead to a systematic deviation of
$0.7$ fm or more. Thus, we re-analyzed the
$\Lambda\Lambda$ invariant mass spectrum given in \cite{Yoon07},
employing our method based on Eq.~(\ref{a}). 
Indeed, since in the $\Lambda\Lambda$ case the FSI can only occur in the 
$^1S_0$ partial wave and, thus, no polarization experiment is required,
our method can be applied also to data like those of $\Lambda\Lambda$ 
production on carbon \cite{Yoon07} from the reaction 
$^{12}C(K^-,K^+  \Lambda\Lambda X)$.
But one has to keep in mind that in reactions on nuclei the energy 
dependence of the production operator is not so well under control.
For example, there could be excitations in the other fragments of the 
reaction process.  
In addition, in the concrete case, the error bars of the 
$\Lambda\Lambda$ invariant mass distribution are quite large, therefore one has to 
expect large uncertainties for the extracted scattering length. Nevertheless, 
in order to demonstrate how the method works we applied it to the data of
Ref.~\cite{Yoon07}, following the procedure for the analysis of experimental 
data described in detail in the Appendix A of \cite{Gasparyan:2003cc}.
We fit the data with the amplitude squared parametrized as 
\begin{eqnarray}
|A(m)|^2=\exp{\left[C_0+\frac{C_1^2}{(m^2-C_2^2)}  \right]} \ ,
\label{param1}
\end{eqnarray} 
multiplied with the phase-space factor, and allowing for a 
finite mass resolution of 2.5 MeV. 
The resulting curve is shown in Fig.~\ref{Yoon} (solid line). 
Then we use this fit to extract the scattering length from the dispersion 
integral with the cut off $\epsilon_{max}=60$ MeV. The result is 
\begin{eqnarray}
a_{\Lambda\Lambda}= -1.2\pm 0.6 \pm 0.4 \ {\rm fm}, 
\label{Result}
\end{eqnarray} 
where the first error is due to the uncertainties in the data 
and the second value is the theoretical error estimated in the 
preceeding subsection. 
 
\begin{table}[h]
\caption{$\Lambda\Lambda$ \,$^1S_0$ scattering lengths ($a_s$) and effective 
range parameters ($r_s$) for various strangeness $S=-2$ interaction potentials (in fm). 
In case of the $\chi$EFT interaction results for the lowest (550 MeV) and 
highest (700 MeV) cut-off value are given, cf. \cite{Polinder07}.
Note that (a) the scattering lengths of the Nijmegen (ESC04)
potential differ significantly depending on whether
they are calculated in particle \cite{Rijken06} or
isospin \cite{Rijken06a} basis, (b) in the potentials 
by Tominaga et al.~\cite{Ueda98} some channel couplings 
are not included. 
}
\vskip 0.2cm
\begin{tabular}[t]{|c|c|c|c|}
\hline
$YY$ interaction & reference & $a_s$ [fm] & $r_s$ [fm] \\
\hline
$\chi$EFT (550) &\cite{Polinder07} & -1.52 & 0.82  \\
$\chi$EFT (700) &\cite{Polinder07} & -1.67 & 0.34 \\
Nijmegen (NSC97a) &\cite{Stoks99} & -0.27 & 15.00 \\
Nijmegen (NSC97f) &\cite{Stoks99} & -0.35 & 14.68 \\
Nijmegen (ESC04a) &\cite{Rijken06a} & -3.804 & 2.42 \\
Nijmegen (ESC04d) &\cite{Rijken06a} & -1.555 & 3.62 \\
Nijmegen (ESC08a'') &\cite{Rijken10} & -0.88 & 4.34 \\
Tominaga (set B) &\cite{Ueda98} & -3.40 & 2.79  \\
Fujiwara (fss2) &\cite{Fujiwara07} & -0.821 & 3.78  \\
Valcarce &\cite{Valcarce10} & -2.54 & -   \\
\hline
\end{tabular}
\label{Table1}
\end{table}

For the ease of comparison we present in Table \ref{Table1} a selection 
of $\Lambda\Lambda$ $^1S_0$ scattering lengths predicted by various 
published $YY$ interaction models.
The large scattering length of the interaction by Tominaga et al.
from 1998 \cite{Ueda98} was still triggered by 
the first experimental information on the ground states of 
${}_{\Lambda\Lambda}^{\;\;\;6}{\rm He}$,
${}_{\Lambda\Lambda}^{\;10}{\rm Be}$, and
${}_{\Lambda\Lambda}^{\;13}{\rm B}$
\cite{Dan63,Prowse:1966nz,Aoki:1991ip}. 
Those experiments suggested a separation energy of 
$\Delta B_{\Lambda\Lambda} = 4-5$ MeV, where the
separation energy is defined as, e.g., 
$\Delta B_{\Lambda\Lambda}=B_{\Lambda\Lambda}
({}^{\;\;\;6}_{\Lambda\Lambda}{\rm He})-2B_{\Lambda}({}^{5}_{\Lambda}{\rm He})$. 
Such a large separation energy could be only described with a rather 
strong $\Lambda\Lambda$ interaction with scattering lengths $a_{\Lambda\Lambda}$ 
in the order of -2 to -3.6 fm \cite{Afnan03,Ueda98} or larger \cite{Filikhin02}. 
  
The analysis of the unambiguously identified ${}_{\Lambda\Lambda}^{\;\;\;6}{\rm He}$
double hypernucleus (Nagara event) \cite{Takahashi:2001nm} yielded the 
much smaller separation energy of $1.01\pm0.20$ MeV. 
Calculations that obtain separation energies in agreement with the new 
experimental value suggest $\Lambda\Lambda$ scattering lengths in the order of 
$-0.7$ to $-1.3$ fm \cite{Filikhin02,Fujiwara07,Rijken06}.
It should be said, however, that so far there are no fully microscopic 
(i.e. six-body) calculations of ${}_{\Lambda\Lambda}^{\;\;\;6}{\rm He}$ 
available that utilize only elementary baryon-baryon ($NN$, $YN$, $YY$)
interactions such as those listed in Table \ref{Table1}.  
All studies are performed either with three-body Faddeev equations applied 
to the cluster model \cite{Filikhin02,Filikhin03,Filikhin04,Hiyama10,Fujiwara04}, 
the Brueckner theory approach \cite{Rijken06,Vidana04},  
or with the stochastical variational method \cite{Nemura05}, 
and rely, at least partly, on effective two-body interactions. 

In this context let us mention that the separation energy for the Nagara event has
been recently re-analysed and is now given as
$\Delta B_{\Lambda\Lambda} = 0.67\pm0.17$ MeV \cite{Nakazawa}. See also
Ref.~\cite{Gal10} for a recent review of the status of ${\Lambda\Lambda}$ hypernuclei. 

\begin{table}[h]
\caption{$\Sigma\Sigma$ and $\Xi N$ 
$S$-wave scattering lengths $a$ and effective range parameters
$r$ for various strangeness $S=-2$ interaction potentials (in fm). 
The subscripts $s$ and $t$ refer to the singlet (\,$^1S_0$) and 
triplet (\,$^3S_1$) states, respectively. In case of the $\chi$EFT
interaction results for the lowest (550 MeV) and highest 
(700 MeV) cut-off value are given, cf. \cite{Polinder07}.
Note that (a) the scattering lengths of the Nijmegen (ESC04)
potential differ significantly depending on whether
they are calculated in particle \cite{Rijken06} or
isospin \cite{Rijken06a} basis, (b) in the potentials 
\cite{Ueda98,Ueda01} some channel couplings are not included. 
}
\vskip 0.2cm
\begin{tabular}[t]{|c|c|c|c|c|c|c|}
\hline
$YY$ interaction & reference & channel & $a_s$ [fm] & $r_s$ [fm] & $a_t$ [fm] & $r_t$ [fm]\\
\hline
$\chi$EFT (550) &\cite{Polinder07} &$\Sigma^+\Sigma^+ $ & -6.23  & 2.17 & - & - \\
$\chi$EFT (700) &\cite{Polinder07} &$\Sigma^+\Sigma^+ $ & -9.27  & 1.88 & - & - \\
Nijmegen (NSC97a) &\cite{Stoks99} &$\Sigma^+\Sigma^+ $ & 10.32  & 1.60 & - & - \\
Nijmegen (NSC97f) &\cite{Stoks99} &$\Sigma^+\Sigma^+ $ &  6.98  & 1.46 & - & - \\
Fujiwara (fss2)&\cite{Fujiwara07} &$\Sigma^+\Sigma^+ $ & -85.3  & 2.34 & - & - \\
Valcarce       &\cite{Valcarce10} &$\Sigma^+\Sigma^+ $ & 0.523  &      & - & - \\
\hline
$\chi$EFT (550) &\cite{Polinder07} &$\Xi^0 n $ & - & - & -0.34 & -5.86\\
$\chi$EFT (700) &\cite{Polinder07} &$\Xi^0 n $ & - & - & -0.15 & 16.3\\
Nijmegen (ESC04a) &\cite{Rijken06a} &\ $\Xi N (I=0) $ \ & - & - & -1.672 & 2.70\\
Nijmegen (ESC04d) &\cite{Rijken06a} &$\Xi N (I=0) $ & - & - &  122.5 & 2.083\\
Nijmegen (ESC08a'') &\cite{Rijken10} &$\Xi N (I=0) $ & - & - &    6.9 & 1.18\\
Tominaga (set B) &\cite{Ueda98} &$\Xi N (I=0) $ & - & - & -0.352 & 17.4\\
Ehime (1.82) &\cite{Ueda01} &$\Xi N (I=0) $ & - & - & -0.43 & 13.0 \\
Valcarce &\cite{Valcarce10} &$\Xi N (I=0) $ & - & - &  0.28 & \\
\hline
$\chi$EFT (550) &\cite{Polinder07} &$\Xi^- n $ &  0.21& -30.7 & 0.02  & 968 \\
$\chi$EFT (700) &\cite{Polinder07} &$\Xi^- n $ &  0.13& -98.5 & 0.03  & 548 \\
Nijmegen (NSC97a) &\cite{Stoks99} &$\Xi^- n $ &  0.46& -6.09 & -0.04 & 634 \\
Nijmegen (NSC97f) &\cite{Stoks99} &$\Xi^- n $ &  0.40& -8.88 & -0.31 & 870 \\
Nijmegen (ESC04a) &\cite{Rijken06a} &$\Xi^- n $ &  0.491& -0.421&       &     \\
Nijmegen (ESC04d) &\cite{Rijken06a} &$\Xi^- n $ &  0.144&  4.670&       &     \\
Nijmegen (ESC08a'') &\cite{Rijken10} &$\Xi^- n $ &  0.58 &  -2.71& 3.49 & 0.60 \\
Tominaga (set B) &\cite{Ueda98} &$\Xi^- n $ &  -0.202& 33.0 & -0.484 & 10.6 \\
Ehime (1.82) &\cite{Ueda01} &$\Xi^- n $ &  -0.27& 20.3 & -0.56 & 9.0 \\
Fujiwara (fss2) &\cite{Fujiwara07} &$\Xi^- n $ &  0.324& -8.93 & -0.207 & 26.2 \\
Valcarce        &\cite{Valcarce10} &$\Xi^0 p $ & -3.32&       & 18.69 & \\
\hline
\end{tabular}
\label{Table2}
\end{table}

\section{The $\Xi N$ and $\Sigma\Sigma$ scattering lengths}
Since the $\Lambda\Lambda$ system is a pure isospin $I=0$ state, 
the $I=1$ $\Xi N$ interaction is also elastic and, thus, permits
a determination of the corresponding scattering length via our
method. The first inelastic channel, $\Lambda\Sigma$, opens at
the excess energy of around 52 MeV and, therefore, should affect
the extraction of the scattering length less than what has
been discussed in the context of the $\Lambda\Lambda$ case above. 
The required $\Xi N$ invariant mass spectrum is accessible 
experimentally in the reaction $K^-d \to K^+\Xi^- n$, for example. 
However, since $\Xi N$ can occur in the $^1S_0$ as well as in
the $^3S_1$ partial wave, one needs data from an experiment
with polarization for a separation of the singlet and triplet 
contributions. The relevant observables are discussed in
Appendix~\ref{Observ}. 
In addition, one has to keep in mind that the reaction 
$K^-d \to K^+\Xi N$ might be dominated by the quasi-elastic
process $K^-N \to K^+ \Xi$, similar to what happened 
for the reaction $\gamma d \to KYN$ \cite{Gasparyan:2007mn} 
with $\gamma N\to KY$. In such
a case the reaction kinematics has to be chosen rather carefully in 
order to suppress the contributions from the quasi-elastic process,
as studied in detail in \cite{Gasparyan:2007mn}, which certainly 
increases the difficulties for a corresponding experiment. 

As a subtlety let us mention that even in the $\Xi^0 n$ system the 
scattering length for $^3S_1$ is real, 
although the amplitude is actually the sum of $I=0$ and $I=1$ states. 
Because of the Pauli principle, the $^3S_1$ partial wave of
the $\Lambda\Lambda$ system is forbidden so that there is
no coupling between the $\Lambda\Lambda$ and $\Xi N$ channels for 
the partial wave in question. 

Experimental information on the $\Xi N$ invariant mass spectrum
is scarce \cite{Goyal80,Saclay82,Godbersen95}. 
Results published in Ref. \cite{Goyal80} for the
$\Xi^- p$ case, obtained from a $K^-d$ bubble-chamber experiment,
suggest an enhancement in the invariant mass distribution - 
but at around 2480 MeV and not near the $\Xi N$ threshold 
which is at around 2255 MeV. In any case, the statistics is too 
low for drawing any conclusions. 
The situation is better for an experiment performed at 
Saclay~\cite{Saclay82} where the missing mass ($MM$) in the reaction 
$K^-d \to K^+ + MM$ at 1.4 GeV/c was studied. The curve 
presented in this publication exhibits a rather smooth behaviour
around the $\Xi N$ threshold which suggests that the $\Xi^- n$
interaction in the $^1S_0$ and/or $^3S_1$ might be fairly
weak. Such a conjecture is actually in line with the results 
of several of the potential models summarized in Table~\ref{Table2}
which predict rather small $\Xi^- n$ scattering lengths. 
We want to emphasize, however, that there are also models with
a fairly strong $\Xi N$ interaction. Noteably the latest Nijmegen 
potential (ESC08a'') produces bound states in the $^3S_1$ partial wave 
of the $I=0$ and $I=1$ channels \cite{Rijken10}. The binding
energies are comparable to that of the deuteron and, accordingly,
sizeable near-threshold enhancements in the corresponding invariant
mass spectrum are to be expected if such bound states indeed
exist in nature. 

Besides the $\Xi N$ system the $I=2$ channel $\Sigma^+\Sigma^+$ is 
potentially interesting too because it is also elastic. 
But due to the charge it cannot be produced with a $K^-$ beam on the deuteron. 
However, the $\Sigma^-\Sigma^-$ system which is likewise $I=2$ 
could be studied, namely in the reaction $K^-d \to K^+\pi^+\Sigma^-\Sigma^-$. 
Also in this case, only the $^1S_0$ partial wave is present so that no polarization 
data are required for a determination of the scattering length. 
Contrary to $\Xi^- n$, here practically all model predictions for 
the $\Sigma^+\Sigma^+$ ($\Sigma^-\Sigma^-$) scattering length are fairly 
large, cf. Table~\ref{Table2}. Details of the application of our method to
cases where the Coulomb interaction is present can be found in 
Ref.~\cite{Gasparyan:2005fk}. 
 
In principle, one can even consider reactions of the type $K^-d \to KK\Xi \Lambda$
and $K^-d \to KK\Xi \Sigma$, which would give access to the strangeness $S=-3$ world.
Potential-model calculations \cite{Rijken06,Fujiwara07} and also predictions 
obtained within the framework of $\chi$EFT \cite{Haidenbauer10} 
suggest that the interaction in some of the channels are strongly attractive 
so that the corresponding scattering lengths could be large. 
Clearly, here it would be desirable to have at least a rough estimation of the
count rates that one can expect in order to judge the feasibility of such experiments. 
Independently of that, also for these reactions we provide and discuss 
the relevant polarization observables that are needed for 
a separation of the singlet and triplet contributions, cf. Appendix~\ref{Observ}. 

\section{Summary}
We reviewed a method that allows one to extract hadronic scattering lengths from production reactions
by studying final-state interactions. In particular, we discussed its applicability to the case 
of baryon-baryon interactions in the strangeness $S=-2$ and $S=-3$ sectors. 
We emphasized the importance of separating different spin states of the interacting particles.
Considering as examples the reactions $K^- d\to K B_1B_2$, $\gamma d\to K_1K_2 B_1B_2$ and 
$K^- d\to K_1K_2 B_1B_2$, we could demonstrate that it is possible to construct polarization 
observables that provide access to spin-singlet and spin-triplet scattering lengths.
In case of the $\Lambda\Lambda$ and $\Sigma^+\Sigma^+$ (or $\Sigma^-\Sigma^-$) interactions
near threshold only the $^1S_0$ partial wave is present due to the Pauli principle and,
thus, no polarization experiments are required for determining the pertinent 
scattering length from the final-state interaction. 
Employing the method to available data on the $\Lambda\Lambda$ invariant mass 
from $^{12}C(K^-,K^+\Lambda\Lambda X)$ \cite{Yoon07}, a $^1S_0$ scattering length of 
$a_{\Lambda\Lambda}=-1.2\pm 0.6$ fm is deduced.
The error given here reflects the accuracy of those data. Thus, it would 
be important to perform experiments with better statistics which could be done,
e.g., at J-PARC \cite{AhnJPARC}.  
This would then allow one to reduce the error on the $\Lambda\Lambda$ scattering length to 
the one of the extraction method, which we estimate to be in the order of $0.3 - 0.4$ fm. 

\acknowledgments{
J.H. acknowledges stimulating discussions with 
J.K. Ahn, A. Gillitzer, D. Grzonka, A. Nogga, J. Schaffner-Bielich, and R. Schumacher. 
}


\appendix 

\section{Spin considerations for the production of baryon-baryon systems with strangeness $S=-2$ and $S=-3$}
\label{Observ}

The technique utilized for the hyperon-nucleon interactions in \cite{Gasparyan:2003cc,Gasparyan:2005fk,Gasparyan:2007mn}
is applicable also in the case of strangeness $S=-2$ and $S=-3$ systems.
The necessary condition that baryon-baryon scattering should be elastic up to
some $m=m_{max}$ is satisfied for the following $S=-2$ channels:
$\Lambda\Lambda$, $\Sigma^+\Sigma^+$, $\Sigma^-\Sigma^-$, $\Xi^0 p$,
$\Xi^- n$, and for the $S=-3$ channels: $\Xi^-\Lambda$,
$\Xi^0\Lambda$, $\Xi^-\Sigma^-$, $\Xi^0\Sigma^+$. 
See Fig.~\ref{fig0} for a graphic overview of the kinematics.
In what follows we are going to consider as examples $K^- d$, $pp$, and $\gamma d$ scattering
in complete analogy with the hyperon-nucleon production reactions studied in \cite{Gasparyan:2007mn}.
The following four types of reactions can be used to produce baryon-baryon states with
strangeness $S=-2$ and $S=-3$: 
\begin{eqnarray}
K^- d\to K B_1B_2&& (K^0 \Lambda\Lambda, \  K^+ \Xi^- n), \\
pp\to K_1K_2 B_1B_2&&  
(K^+K^+ \Lambda\Lambda, \ K^+K^0 \Xi^0 p, \ K^0K^0 \Sigma^+\Sigma^+), \\
\gamma d\to K_1K_2 B_1B_2&&  
(K^+K^0 \Lambda\Lambda, \ K^0K^0 \Xi^0 p, \ K^+K^+ \Xi^- n), \\
K^- d\to K_1K_2 B_1B_2&&   
(K^0K^0 \Xi^0\Lambda, \ K^+K^0 \Xi^-\Lambda, \ K^+K^+ \Xi^-\Sigma^-).
\end{eqnarray}

Now we come to the question of separating the spin-triplet and spin-singlet states in
the baryon-baryon system. As in the case of the hyperon-nucleon interaction it is
sufficient to consider reactions with polarized initial particles.
We start from the general form for the reaction amplitude in the center-of-mass (CM) 
system for the three processes
\begin{eqnarray}
{\cal M}_{K^- d\to K B_1B_2}&=&a_1^s(\vec \epsilon_d \times \hat p)\cdot \vec k+a_2^t (\vec \epsilon_d\cdot \vec S')
+a_3^t (\vec \epsilon_d\cdot \hat p) (\vec S'\cdot\vec k)
+a_4^t (\vec \epsilon_d\cdot \vec k) (\vec S'\cdot\hat p) \, ,\nonumber\\ 
{\cal M}_{pp\to K_1K_2 B_1B_2}&=&b_1^s+b_2^s( \hat p\times \hat k)\cdot \vec S+b_3^t(\hat p\cdot \hat k)( \hat p\times \hat k)\cdot \vec S'+
(b_4^t \hat p_i   \hat k_j+b_5^t \hat p_j   \hat k_i)S_i S'_j\, , \nonumber\\
{\cal M}_{\gamma d\to K_1K_2 B_1B_2}&=&c_1^s(\vec \epsilon_\gamma \times \vec \epsilon_d) \cdot \hat p+
c_2^t(\vec \epsilon_\gamma \cdot \vec \epsilon_d)(\vec S'\cdot \hat p)
+c_3^t(\vec \epsilon_\gamma \cdot \vec S')(\vec \epsilon_d\cdot \hat p)\, , \nonumber\\
 {\cal M}_{K^- d\to K_1K_2 B_1B_2}&=&d_1^s(\vec \epsilon_d \cdot \hat p)
+d_2^t (\vec \epsilon_d\times \vec S')\cdot \hat p\, .
\label{amplitude}
\end{eqnarray}
Here $a$, $b$, $c$, and $d$ are some functions of $s$ and of $m$ (and, in general, of further 
invariants that are required to specify the kinematics of the reaction), 
where their upper indices
indicate whether they correspond to spin-singlet ($s$) or spin-triplet ($t$) amplitude. 
The polarization vectors of the deuteron and photon are denoted by 
$\vec \epsilon_d$ and $\vec \epsilon_\gamma$, respectively. 
The spin vectors $\vec S$, $\vec S'$ are used for the spin-triplet initial and 
final states, respectively. For the last two reactions 
we assume the momenta of the final kaons to be either aligned or 
anti-aligned with the direction of the initial 
CM momentum $\hat p$. This leads to a significant simplification allowing one to separate 
different spin states. For the reaction $K^- d\to K B_1B_2$ such a restriction is not necessary 
and the momentum of the emitted kaon is denoted by $\vec k$. 
For the reaction $pp\to K_1K_2 B_1B_2$ we assume for simplicity that both emitted kaons 
go into the same direction $\hat k$.
It is convenient to introduce the following set of polarization observables
\begin{eqnarray}
 O_1&=&(1-\sqrt{2}T^0_{20})\frac{d\sigma_0}{dm^2dt}\, , \
 O_2=(2+\sqrt{2}T^0_{20})\frac{d\sigma_0}{dm^2dt}\, , \ 
 O_3=T^0_{10}\frac{d\sigma_0}{dm^2dt}\, ,\nonumber\\
O_4&=&\left(2+\sqrt{2} T_{20}^0+\sqrt{3} (T_{22}^l+T_{2-2}^l)\right)\frac{d\sigma_0}{dm^2dt}
=\sqrt{3}(-\sqrt{2} T_{10}^c+ \left(T_{22}^l+T_{2-2}^l)\right)\frac{d\sigma_0}{dm^2dt}\, ,\nonumber\\
O_5&=&A_{0y}\frac{d\sigma_0}{dm^2dt}\ , \ O_6=(1+A_{yy})\frac{d\sigma_0}{dm^2dt} \, ,
\end{eqnarray}
where the various $T$'s for the $\gamma d$ initial state are defined in \cite{Gasparyan:2007mn},
and $\frac{d\sigma_0}{dm^2dt}$ is the unpolarized differential cross section.
$T^0_{20}$ and $T^0_{10}$ have the same definition also for the $K^- d$ initial state,
the only difference being the absence of the summation over the photon polarizations.
The observable $O_1$ selects the amplitudes with longitudinal target polarization 
$\vec \epsilon_d\parallel\hat p$, whereas
$O_2$, $O_3$, $O_4$ select the amplitudes with $\vec \epsilon_d\perp\hat p$. 
In addition $O_4$ contain only that part of
the amplitude which is antisymmetric with respect to an interchange of $\vec \epsilon_d$ and $\vec \epsilon_\gamma$.
The observables $O_5$ and $O_6$ correspond to the proton-proton induced reaction. 
Here $A_{0y}$ is the analyzing power and $A_{yy}$ is the spin correlation 
coefficient for polarized beam and target \cite{Hanhart,Gasparyan:2003cc}, 
and $y$ is the direction perpendicular to the reaction plane.

Now inspecting the structure of the reaction amplitudes (\ref{amplitude}) we can identify the 
observables that allows one to separate a particular spin state:
For the reaction $K^- d\to K B_1B_2$ the triplet final state can be singled out by the observable $O_1$
for any direction of the emitted kaon, or one can measure the unpolarized differential cross section
for $\vec k \parallel\hat p$. The spin singlet state cannot be separated.
For the reaction $\gamma d\to K_1K_2 B_1B_2$ the triplet final state can be separated by measuring $O_1$.
The observable $O_4$ provides access to the spin-singlet amplitude.
For the reaction $K^- d\to K_1K_2 B_1B_2$ the observable $O_1$ separates spin-singlet contribution,
whereas $O_2$ and $O_3$ separate spin-triplet state.
For the reaction $pp\to K_1K_2 B_1B_2$ the observable $O_6$ separates the spin-triplet contribution,
whereas $O_5$ is proportional to 
$\sin{\theta}\,\Im \{b_1^s b_2^{s *} + b_4^t b_5^{t *}\cos{\theta} \}$ 
with $\cos{\theta}=\hat p\cdot\hat k$. Since the $b_i$'s are
even functions of $\cos{\theta}$ (due to parity conservation), after the integration 
of $O_5$ over an angular region symmetric with respect to $\theta=\frac{\pi}{2}$ 
only spin-singlet amplitudes survive.

\end{document}